\documentclass{llncs}

\usepackage[T1]{fontenc}
\usepackage{url}

\usepackage{stmaryrd}
\usepackage{amssymb}
\usepackage[all]{xy}
\usepackage{color}

\renewcommand{\tt}{\ttfamily}
\newcommand{\codefont}{\tt\small}
\newcommand{\code}[1]{\mbox{\codefont{#1}}}
\newcommand{\ccode}[1]{``\code{#1}''}
\newcommand{\us}{\raise-.8ex\hbox{-}}
\newcommand{\xtilde}{\!\raise-.75ex\hbox{\char`\~}} 
\newcommand{\setfun}[1]{#1\ensuremath{_{\cal S}}}
\newcommand{\seq}{\mathop{\code{==}}} 
\newcommand{\sid}{\mathop{\code{===}}} 
\newcommand{\ceq}{\mathop{\code{=:=}}} 

\usepackage{listings}
\lstnewenvironment{curry}[1][]
  {\lstset{literate={->}{{$\rightarrow{}\!\!\!$}}3,#1}}{}
\newcommand{\listline}{\vrule width0pt depth1.5ex}

\newcommand{\equprogram}[1]{%
\def\separator{0.5ex}%
\frenchspacing%
\refstepcounter{equation}%
\par\vspace\separator\hspace{-0.8em}%
$\vcenter{\codefont\noindent{#1}}$%
\raisebox{-0.0ex}{\kern-0.5em\llap{\rm (\theequation)}}
\par\vspace\separator\noindent\kern-.0em%
}
\begin{document}
\pagestyle{plain} 
\sloppy

\title{Adding \texttt{Data} to Curry}

\author{
Michael Hanus
\kern1em
Finn Teegen
}
\institute{
Institut f\"ur Informatik, CAU Kiel, 24098 Kiel, Germany\\
\email{\{mh,fte\}@informatik.uni-kiel.de}
}

\maketitle

\begin{abstract}
Functional logic languages can solve equations over user-defined
data and functions.
Thus, the definition of an appropriate meaning of equality
has a long history in these languages, ranging from
reflexive equality in early equational logic languages
to strict equality in contemporary functional logic languages like Curry.
With the introduction of type classes, where the equality operation
\ccode{==} is overloaded and user-defined, the meaning became
more complex.
Moreover, logic variables appearing in equations
require a different typing than pattern variables,
since the latter might be instantiated with functional values or
non-terminating operations.
In this paper, we present a solution to these problems
by introducing a new type class \code{Data} which is associated
with specific algebraic data types, logic variables,
and strict equality.
We discuss the ideas of this class and its implications
on various concepts of Curry, like unification, functional patterns,
and program optimization.
\end{abstract}

\section{Introduction}
\label{sec:intro}

The amalgamation of the main declarative programming paradigms,
namely functional and logic programming, has a long history.
The advantages of such integrated functional logic languages
are manifold. One can use the features of functional programming
(e.g., powerful type systems, higher-order functions, lazy evaluation)
and logic programming (e.g., non-deterministic search,
computing with partial information) in a single language
which also leads to new design patterns
\cite{AntoyHanus02FLOPS,AntoyHanus11WFLP}.
Compared to logic programming, computations can be more efficient
due to the use of optimal evaluation strategies \cite{AntoyEchahedHanus00JACM}.

Early approaches to integrating functional and logic programming
(see \cite{DeGrootLindstrom86} for a good collection of these proposals)
used equational logic programming \cite{GoguenMeseguer86,ODonnell98}
as a unifying framework.
From a logic programming point of view,
equational logic programming extends the meaning of the
standard equality predicate \ccode{=}
by taking user-defined functions into account before checking
the equality of both sides of an equation.
Hence, both sides are evaluated before they are unified.
If the definition of evaluable functions are considered as axioms
for an equational theory, this process is also known as
\emph{E-unification} \cite{GallierSnyder89}.
In order to use logic programming techniques (computing with partial
information) also for the evaluation of user-defined functions,
one can use narrowing instead of reduction \cite{Reddy85},
i.e., replace pattern matching by unification when
a function call should be reduced.
In this way, functional logic languages based on narrowing
can be used to \emph{solve equations}.

\begin{example}\label{ex:peano-add}
Consider the following definition of Peano numbers
and their addition (in Haskell \cite{PeytonJones03Haskell} syntax):
\begin{curry}
data Nat = Z | S Nat$\listline$
add :: Nat -> Nat -> Nat
add Z     n = n
add (S m) n = S (add m n)
\end{curry}
In the functional language Haskell, we can only compute the value
of expressions, e.g.,
\begin{curry}
> add (S Z) (S Z)
S (S Z)
\end{curry}
However, if we interpret these definitions as
a program written in the (narrowing-based) functional logic language
Curry \cite{Hanus97POPL,Hanus16Curry},
we can also solve the equation
\begin{curry}
> add x (S Z) =:= S (S Z)   where x free
{x$\,$=$\,$S Z} True
\end{curry}
Here, \ccode{=:=} denotes equality w.r.t.\ user-defined operations
(see below for more details) and \code{x} is declared as a
free (logic) variable which is bound to \code{S$\;$Z} in order
to evaluate the equation to \code{True}.
\end{example}
For the practical applicability of functional logic languages,
it is important to reduce the computation space
by using specific evaluation strategies.
Thus, much work in this area has been devoted to develop
appropriate narrowing strategies (see \cite{Hanus94JLP}
for an early account of this research).
In order to provide the advantages of lazy evaluation
used in Haskell, e.g., optimal evaluation \cite{HuetLevy91}
and modularity \cite{Hughes90},
later research concentrated on demand-driven strategies.
Needed narrowing \cite{AntoyEchahedHanus00JACM} is an optimal
strategy \cite{Antoy97ALP} and, thus, the basis of the
language Curry.

Demand-driven evaluation strategies, like Haskell's lazy evaluation
or Curry's needed narrowing, can deal with non-terminating operations
that compute infinite data structures \cite{Hughes90}.
However, this could be in conflict with the equation solving
capabilities of functional logic languages discussed above.
Standard equality in the mathematical sense is required to
be reflexive, i.e., $x = x$ should always hold \cite{ODonnell98}.
Now consider two operations to compute infinite lists of Peano numbers:
\begin{curry}
f1 :: Nat -> [Nat]
f1 n = n : f1 (S n)$\listline$
f2 :: Nat -> [Nat]
f2 n = n : S n : f2 (S (S n))
\end{curry}
By reflexivity, $\code{f1$\;$Z} = \code{f1$\;$Z}$ should hold.
This means that the infinite lists of all Peano numbers are equal.
As a consequence, $\code{f1$\;$Z} = \code{f2$\;$Z}$ should also hold,
but it is unclear to verify it during run time.
In early equational logic programming,
equations are solved by narrowing both sides to normal forms
and unifying these normal forms.
However, this does not work here since \code{f1$\;$Z} and \code{f1$\;$Z}\
have no normal form.
Thus, reflexivity is not a feasible property of equations to be evaluated
(more details including issues about semantics are discussed in
\cite{GiovannettiLeviMoisoPalamidessi91,MorenoRodriguez92}).

Therefore, contemporary languages interpret equations to be evaluated
as \emph{strict equality},
denoted by \ccode{=:=} in Curry:
$e_1 \ceq e_2$ is satisfied iff $e_1$ and $e_2$ are reducible
to a same ground constructor term, i.e., an expression
without variables and defined functions.
In particular, soundness, completeness, and optimality results
are stated w.r.t.\ strict equality \cite{AntoyEchahedHanus00JACM}.
As a consequence, \code{f1$\;$Z =:= f1$\;$Z} does not hold
so that it is not a defect that this equation cannot be solved.

Note that Haskell also offers the operation \ccode{==}
intended to compare expressions.
Although standard textbooks on Haskell define this operation
as ``equality'' \cite{Bird98,Hutton16,Thompson99Haskell},
its actual implementation can be different since, as a member of the
type class \code{Eq}, it can be defined with a behavior
different than equality on concrete type instances.
Actually, the documentation of the type class \code{Eq}\footnote{%
\url{http://hackage.haskell.org/package/base-4.12.0.0/docs/Data-Eq.html}}
denotes \ccode{==} as ``equality'' but also contains the remark:
``\code{==} is customarily expected to implement an equivalence
relationship where two values comparing equal are indistinguishable
by ``public'' functions.''
Thus, it is intended that $e_1 \seq e_2$ evaluates to \code{True}
even if $e_1$ and $e_2$ have not the same but only equivalent values.
On the other hand, the documentation requires that the
reflexivity property
\begin{curry}
x == x = True
\end{curry}
holds for any implementation, but this is not true even for the
standard integer equality
(choose \ccode{last [1..] :: Int} for \code{x}).

This discussion shows that the precise treatment of equality,
which is essential for functional logic languages,
might have some pitfalls when type classes are used.
As long as \ccode{==} is defined in the standard way
(by the use of \ccode{deriving Eq}), \ccode{==} conforms with
strict equality.
With the introduction of type classes to Curry,
one has to be more careful.
For instance, consider the ``classical'' functional logic definition
of the operation \code{last} to compute the last element
of a list by exploiting list concatenation (\ccode{++})
and equation solving \cite{Hanus94JLP,Hanus13}:\label{sec:last}
\begin{curry}
last xs | _ ++ [e] == xs = e
  where e free
\end{curry}
If \ccode{==} denotes equivalence rather than strict equality,
\code{last} might not return the last element of a list
but one (or more than one) value which is equivalent to the
last element.

In this paper, we propose a solution to these problems
by distinguishing between strict equality and equivalence.
For this purpose, we propose a new type class \code{Data}
which is associated with specific algebraic data types.
We will see that this type class can also be used
for a better characterization of the meaning of logic variables
and the Curry's unification operator \ccode{=:=}.

This paper is structured as follows.
In the next section, we review some aspects of functional logic programming
and Curry.
After motivating the problem this paper tackles in Sect.~\ref{sec:equivalence},
we propose in Sect.~\ref{sec:data} a new standard type class for Curry,
namely \code{Data}, as a solution to the problem.
In Sect.~\ref{sec:freevars}, Sect.~\ref{sec:eqopt}, and Sect.~\ref{sec:funpats},
we discuss how the proposed \code{Data} type class affects logic variables,
optimization of equality constraints, and non-left-linear rules and
functional patterns, respectively.
Finally, Sect.~\ref{sec:related} discusses related work
before we conclude in Sect.~\ref{sec:concl}.

\section{Functional Logic Programming and Curry}
\label{sec:flp}

We briefly review some aspects of functional logic programming
and Curry that are necessary to understand the contents of this paper.
More details can be found in surveys on
functional logic programming \cite{AntoyHanus10CACM,Hanus13}
and in the language report \cite{Hanus16Curry}.

Curry is a declarative multi-paradigm language intended to
combine the most important features from functional and logic programming.
The syntax of Curry is close to Haskell \cite{PeytonJones03Haskell}
but also allows \emph{free} (\emph{logic}) \emph{variables}
in conditions and right-hand sides of rules.
Thus, \emph{expressions} in Curry programs contain
\emph{operations} (defined functions),
\emph{constructors} (introduced in data type declarations),
and \emph{variables} (arguments of operations or free variables).
Function calls with free variables are evaluated by a possibly
non-deterministic instantiation of demanded arguments
\cite{AntoyEchahedHanus00JACM}.
This corresponds to narrowing \cite{Reddy85},
but Curry narrows with possibly non-most-general unifiers
to ensure the optimality of computations \cite{AntoyEchahedHanus00JACM}.
In contrast to Haskell, rules with overlapping left-hand sides
are non-deterministically (rather than sequentially) applied.

\begin{example}\label{ex:concdup}
The following simple program shows the functional and logic features
of Curry. It defines the well-known list concatenation
and an operation that returns
some element of a list having at least two occurrences:
\begin{curry}
(++) :: [a] -> [a] -> [a]
[]     ++ ys = ys
(x:xs) ++ ys = x : (xs ++ ys)$\listline$
someDup :: [a] -> a
someDup xs | xs =:= _$\,$++$\,$[x]$\,$++$\,$_$\,$++$\,$[x]$\,$++$\,$_ = x
  where x free
\end{curry}
Since \ccode{++} can be called with free variables in arguments,
the condition in the rule of \code{someDup}
is solved by instantiating \code{x} and
the anonymous free variables \ccode{\us} to appropriate values
before reducing the function calls.
As already mentioned in the introduction,
\ccode{=:=} denotes strict equality, i.e., the condition
of \code{someDup} is satisfied if both sides are reduced to
a same ground constructor term.
In order to avoid the enumeration of useless values,
\ccode{=:=} is implemented as unification:\label{sec:impl-unif}
if \code{y} and \code{z} are free (unbound) variables,
$\code{y} \ceq \code{z}$ is evaluated (to \code{True})
by binding \code{y} and \code{z}
(or vice versa) instead of non-deterministically binding
\code{y} and \code{z} to identical ground constructor terms.
This can be interpreted as an optimized implementation
by delaying the bindings to ground constructor terms
\cite{AntoyHanus17FAoC}.
Due to this implementation, \ccode{=:=} is also called
an \emph{equational constraint} (rather than Boolean equality).
\end{example}
We already used the logic programming features of Curry
in the definition of \code{last} shown in Sect.~\ref{sec:last}.
In contrast to \code{last},
\code{someDup} is a \emph{non-deterministic operation}
since it could yield more than one result for a given argument,
e.g., the evaluation of \code{someDup$\,$[1,2,2,1]} yields the values
\code{1} and \code{2}.
Non-deterministic operations, which can formally be
interpreted as mappings from values into sets of values \cite{GonzalezEtAl99},
are an important feature
of contemporary functional logic languages.
Hence, Curry has also a predefined \emph{choice} operation:
\begin{curry}
x ? _  =  x
_ ? y  =  y
\end{curry}
Thus, the expression \ccode{0$~$?$~$1} evaluates to \code{0} and \code{1}
with the value non-deterministically chosen.

\section{Equality vs.\ Equivalence}
\label{sec:equivalence}

Type classes are an important feature to express
ad-hoc polymorphism in a structured manner \cite{WadlerBlott89}.
In the context of Curry, it is also useful to restrict
the application of some operations to unintended expressions.
For instance, in the definition of Curry without type classes
\cite{Hanus16Curry}, the type of the unification operator
is defined as
\begin{curry}
(=:=) :: a -> a -> Bool
\end{curry}
This implies that we could unify values of any type,
including defined functions.
However, the meaning of equality on functions is not well defined.
The Curry implementation PAKCS \cite{Hanus18PAKCS},
which compiles Curry programs into Prolog programs,
uses an intensional meaning, i.e., functions are equal
if they have the same name. This means that PAKCS evaluates
\begin{curry}
not =:= not
\end{curry}
to \code{True} but it fails on
\begin{curry}
not =:= (\x -> not x)
\end{curry}
(since the lambda abstraction will be lifted into a new top-level function).
Moreover, the Curry implementation KiCS2
\cite{BrasselHanusPeemoellerReck11},
which compiles Curry programs into Haskell programs,
produces an internal error for these expressions.

It would be preferable to forbid the application of \ccode{=:=}
to functional values at compile time.
This is similar to the requirement on Haskell's operator \ccode{==}.
Haskell uses the type class \code{Eq}
in order to express that \ccode{==} is not parametric polymorphic
but overloaded for some (but not all) types.
The type class \code{Eq} contains two operations
(we omit the default implementations):
\begin{curry}
class Eq a where
  (==) :: a -> a -> Bool
  (/=) :: a -> a -> Bool
\end{curry}
Hence, the operator \ccode{==} cannot be applied to any type but only
to types defining instances of this class.
We can use this operator to check whether
an element occurs in a list:
\begin{curry}
elem :: Eq a => a -> [a] -> Bool
elem _ []     = False
elem x (y:ys) = x==y || elem x ys
\end{curry}
Although type classes express type restrictions in an elegant manner,
they might also cause unexpected behaviors if they are not carefully used.
For instance, we can define a data type for values indexed by a unique
number:
\begin{curry}
data IVal a = IVal Int a
\end{curry}
Since the index is assumed to be unique, we define the comparison
of index values by just comparing the indices:
\begin{curry}
instance Eq a => Eq (IVal a) where
  IVal i1 _ == IVal i2 _  =  i1 == i2
\end{curry}
With this definition, the operation \code{elem} defined above
could yield surprising results:
\begin{curry}
> elem (IVal 1 'b') [IVal 1 'a']
True
\end{curry}
This is not intended since the element (first argument)
does not occur in the list.
Actually, the Haskell documentation\footnote{%
\url{http://hackage.haskell.org/package/base-4.12.0.0/docs/Prelude.html}}
about \code{elem} contains the explanation
``Does the element occur in the structure?''
which ignores the fact that some instances of \code{Eq}
are only equivalences rather than identities.

This unusual behavior could also influence logic-oriented
computations in a surprising manner.
If the operation \code{last} is defined as shown in Sect.~\ref{sec:last},
we obtain the following answer when computing the last element
of a given \code{IVal} list (here, \ccode{\us} denotes a logical variable of type \code{Char}):
\begin{curry}
> last [IVal 1 'a']
IVal 1 _
\end{curry}
Hence, instead of the last element, we get a rather general
representation of it.

The next section presents our proposal to solve these problems.

\section{\texttt{Data}}
\label{sec:data}

As discussed above, type classes are an elegant way
to express type restrictions.
On the other hand, it is not a good idea to allow 
user-defined instance definitions of important operations
like strict equality.
Therefore, we propose the introduction of a specific type class
where only standard instances can be derived so that all
instances satisfy the intended meaning.
This type class is called \code{Data} and has the following
definition:
\begin{curry}
class Data a where
  aValue :: a
  (===)  :: a -> a -> Bool
\end{curry}
Thus, any instance of this class provides two operations:
\begin{itemize}
\item
The non-deterministic operation \code{aValue} returns some value,
i.e., the complete evaluation of \code{aValue} yields
all values of type \code{a}.
\item
The operation \ccode{===} implements the standard equality on values,
i.e., it returns \code{True} or \code{False} depending on whether
the argument values are identical or not.
\end{itemize}
The following definition specifies how to automatically derive
a \code{Data} instance for any algebraic datatype.
\begin{definition}\label{def:derive}
If $T$ is an algebraic datatype declared by
\begin{curry}
data $T$ $a_1$ $\ldots$ $a_k$ = $C_1$ $b_{11}$ $\ldots$ $b_{1k_1}$ | $\ldots$ | $C_n$ $b_{n1}$ $\ldots$ $b_{nk_n}$
\end{curry}
the standard derived \code{Data} instance has the following form:
\begin{curry}
instance $cx$ => Data ($T$ $a_1$ $\ldots$ $a_k$) where
  aValue = $C_1$ aValue $\ldots$ aValue ? $\ldots$ ? $C_n$ aValue $\ldots$ aValue$\listline$
  $C_1$ $x_1$ $\ldots$ $x_{k_1}$ === $C_1$ $y_1$ $\ldots$ $y_{k_1}$ = $x_1$ === $y_1$ && $\ldots$ && $x_{k_1}$ === $y_{k_1}$
  $\vdots$
  $C_n$ $x_1$ $\ldots$ $x_{k_n}$ === $C_n$ $y_1$ $\ldots$ $y_{k_n}$ = $x_1$ === $y_1$ && $\ldots$ && $x_{k_n}$ === $y_{k_n}$
  $C_i$ _ $\ldots$ _ === $C_j$ _ $\ldots$ _ = False  $\forall i,j \in \{1,\ldots,n\}~ \mathit{with}~ i \neq j$
\end{curry}
In the instance declaration above, the context $cx$ consists of \code{Data}
constraints ensuring that \code{Data $b_{ij}$} holds
for each type $b_{ij}$ with $i \in \{1, \ldots, n\}$
and $j \in \{1, \ldots, k_i\}$.
\end{definition}
\begin{example}
For the type of Peano numbers (see Ex.~\ref{ex:peano-add}),
the \code{Data} instance can be defined as follows:
\begin{curry}
instance Data Nat where
  aValue = Z  ?  S aValue$\listline$
  Z   === Z   = True
  S m === S n = m === n
  Z   === S _ = False
  S _ === Z   = False
\end{curry}
A \code{Data} instance for lists requires a \code{Data} instance
for its elements:
\begin{curry}
instance Data a => Data [a] where
  aValue = []  ?  aValue : aValue$\listline$
  []     === []     = True
  (x:xs) === (y:ys) = x === y && xs === ys
  []     === (_:_)  = False
  (_:_)  === []     = False
\end{curry}
\end{example}
The operation \code{aValue} is useful when a value of some data type
should be guessed, e.g., for testing \cite{Hanus16LOPSTR}.
The obvious relation to logic variables will be discussed later.

The definition of \ccode{===} is identical to \ccode{==}
if the definition of the latter is automatically derived
(by a \ccode{deriving Eq} clause).
As discussed above, it is also possible to define other
instances of \code{Eq} that leads to unintended results.
To ensure that \ccode{===} always denotes equality on values,
\emph{it is not allowed to define explicit \code{Data} instances}
as shown above.
Such instances can only be generated by
adding a \ccode{deriving Data} clause to a data definition.
Note that an instance derivation requires that
all arguments of all data constructors have \code{Data} instances.
In particular, if some argument has a functional type, e.g.,
\begin{curry}
data IntRel = IntRel (Int -> Bool)
\end{curry}
then a \code{Data} instance can not be derived.

For ease of use, one could always derive \code{Data} instances
for data declarations whenever it is possible (i.e., functional values
do not occur in arguments), or provide a language option
to turn this behavior on or off.

With the introduction of the class \code{Data},
we can specify a more precise type to
Curry's strict equality operation \ccode{=:=}.
As discussed in \cite{AntoyHanus17FAoC},
the meaning of \ccode{=:=} is the ``positive'' part of \ccode{===},
i.e., its semantics can be defined by
\equprogram{\label{def:ceq}%
x =:= y ~=~ solve (x === y)
}
where \code{solve} is an operator that enforces positive evaluations
for Boolean expressions:
\equprogram{\label{def:solve}%
solve True = True
}
Since expressions of the form $e_1 \ceq e_2$ might return \code{True}
but never \code{False}, \ccode{=:=} can be implemented
by unification, as already discussed in Sect.~\ref{sec:impl-unif}.
Such an optimized implementation is justified by the
definition (\ref{def:ceq}) above.
However, if the semantics of \ccode{=:=} is defined by
\equprogram{\label{def:ceq-wrong}%
x =:= y ~=~ solve (x == y)
}
as suggested before the introduction of type classes to Curry
\cite{AntoyHanus14},
an implementation of \ccode{=:=} by unification
would not be correct since unification might put stronger requirements
on expressions to be compared than actually defined by \code{Eq} instances.

As a spin-off of definition (\ref{def:ceq}),
we obtain a more restricted type of \ccode{=:=}:
\equprogram{\label{def:ceq-type}%
(=:=) :: Data a => a $\to$ a $\to$ Bool
}
This avoids the problems with the application of \ccode{=:=}
to functional values sketched at the beginning
of Sect.~\ref{sec:equivalence}.

\section{Logic Variables}
\label{sec:freevars}

When a function call with free variables in arguments is evaluated
by narrowing, the free variables are instantiated to values so that
the function call becomes reducible.
Conceptually, a free variable denotes possible values
so that a computation can pick one in order to proceed.
With the definition of the type class \code{Data}
and the non-deterministic operation \code{aValue},
we make the notion of ``possible value'' explicit.
Actually, it has been shown
that non-deterministic operations and logic variables
have the same expressive power
\cite{AntoyHanus06ICLP,deDiosCastroLopezFraguas07}
since one can replace logic variables occurring in a
functional logic program by non-deterministic value generators.

\begin{example}
Consider the addition on Peano numbers
shown in Ex.~\ref{ex:peano-add} which is exploited
to define subtraction:
\begin{curry}
sub :: Nat -> Nat -> Nat
sub x y | add y z === x = z
  where z free
\end{curry}
We can replace the logic variable \code{z} by a value generator:
\begin{curry}
sub x y | add y z === x = z
  where z = aValue
\end{curry}
\end{example}
The equivalence of logic variables and non-deterministic value generators
can be exploited when Curry is implemented
by translation into a target language without support for
non-determinism and logic variables.
For instance, KiCS2 \cite{BrasselHanusPeemoellerReck11}
compiles Curry into Haskell by adding only a mechanism
to handle non-deterministic computations.
Therefore, KiCS2 is able to evaluate a logic variable
to all its values.
Thus, KiCS2 could exploit this fact by using the following alternative definition for \code{aValue}:
\equprogram{\label{def:avalue}%
aValue = \us{}
}
This equivalence also sheds some new light on the type
of logic variables.
Currently, logic variables without any constraints on their types
are considered to have a polymorphic type.
For instance, the inferred type of \code{aValue} as defined in
(\ref{def:avalue}) is
\begin{curry}
aValue :: a
\end{curry}
However, this type does not really describe the intent
of this operation, since \code{aValue} does not yield
functional values.
For instance, consider the definition
\begin{curry}
f x = y where y free
\end{curry}
The type currently inferred is
\begin{curry}
f :: a -> b
\end{curry}
However, it is meaningless to use the result of some application
of \code{f} in contexts where a function is required.
For instance, the evaluation of the expression
\equprogram{\label{def:map-f}%
map (f True) [0,1]
}
suspends in PAKCS and produces a run-time error in KiCS2
(very similar to the examples described at the beginning
of Sect.~\ref{sec:equivalence}).
Furthermore, the inferred type of the definition
\begin{curry}
g x = g x
\end{curry}
is
\begin{curry}
g :: a -> b
\end{curry}
Thus, it looks very similar to the type of \code{f}
although \code{g} has a quite different meaning:
in contrast to \code{f},
an application of \code{g} never returns
a value.

All these problems can be avoided by a simple fix:
logic variables are considered as equivalent
to the operation \code{aValue} of type class \code{Data}
so that a logic variable without any constraints on its type has
type \code{a} where \code{a} is constrained with the type class context
\code{Data$\;$a}.
With this change, the inferred type of \code{f} is
\begin{curry}
f :: Data b => a -> b
\end{curry}
As a consequence, expression (\ref{def:map-f}) will be rejected
by the type checker since functions have no \code{Data} instance.

\section{Equality Optimization}
\label{sec:eqopt}

Choosing the appropriate kind of equality might not be obvious
to the programmer.
The difference between identity and equivalence is
semantically relevant so that the decision between
\ccode{===} and \ccode{==} is not avoidable.
However, \ccode{=:=} can be considered as an optimization
of \ccode{===} so that it is not obvious when it should be applied.
In order to simplify this situation,
it has been argued in \cite{AntoyHanus14,AntoyHanus17FAoC}
that the programmer should always use strict equality
(i.e., \ccode{===}) and the selection of \ccode{=:=}
should be done by an optimization tool.
This tool analyzes the required values of Boolean expressions.
If an application of strict equality requires only
the result value \code{True}, e.g., in guards of conditional rules
or in arguments of \code{solve}, see (\ref{def:solve}),
then one can safely replace the equality operator by
the unification operator \ccode{=:=}
(see \cite{AntoyHanus17FAoC} for details).
For instance, if \code{last} is defined by
\begin{curry}
last xs | _ ++ [e] === xs
        = e  where e free
\end{curry}
then it can be transformed into
\begin{curry}
last xs | _ ++ [e] =:= xs
        = e  where e free
\end{curry}
As shown in \cite{AntoyHanus17FAoC}, this transformation can have
a big impact on the execution time.

Up to now, this tool (which is part of the compilation chain
of Curry systems) considered the optimization of calls to \ccode{==}.
Since this might lead to incompleteness, as discussed above,
it has to consider calls to \ccode{===} when the type class
\code{Data} is introduced.
However, for backward compatibility and better optimizations,
one can extend the optimizer also to calls of the form
$e_1 \seq e_2$: if the types of the arguments $e_1,e_2$
are monomorphic and the \code{Eq} instances of these types
are derived with the default scheme (by \code{deriving} annotations),
the semantics of \ccode{==} is identical to the semantics of \ccode{===}
so that one can replace $e_1 \seq e_2$ by $e_1 \sid e_2$
and apply the optimization sketched above.

\section{Non-Left-Linear Rules and Functional Patterns}
\label{sec:funpats}

The proposed introduction of the type class \code{Data}
together with the adjusted type of the unification operator \ccode{=:=}
has also some influence on language constructs
where unification is implicitly used.
We discuss this in more detail in this section.

In contrast to Haskell, Curry allows non-left-linear rules,
i.e., defining rules with multiple occurrences of a
variable in the patterns of the left-hand side.
For instance, this function definition is valid in Curry:
\begin{curry}
f x x = x
\end{curry}
Multiple occurrences of variables in the left-hand side
are considered as an abbreviation for equational constraints
between these occurrences \cite{Hanus16Curry}, i.e.,
the definition above is expanded to
\begin{curry}
f x y | x =:= y = x
\end{curry}
This feature of Curry is motivated by logic programming
where multiple variable occurrences in rule heads are also solved
by unification.
However, in Curry the situation is a bit more complex
due to the inclusion of functions and infinite data structures.
As a matter of fact, our refined type of \ccode{=:=}
makes the status of non-left-linear rules clearer.
According to the type shown in (\ref{def:ceq-type}),
the type inferred for the definition above is
\begin{curry}
f :: Data a => a -> a -> a
\end{curry}
Hence, \code{f} can not be called with functional values as arguments.
This even increases the compatibility with logic programming
where unification is applied to Herbrand terms, i.e.,
algebraic data.

Another feature of Curry, where equational constraints are
implicitly used, are functional patterns.
\emph{Functional patterns} are proposed in \cite{AntoyHanus05LOPSTR}
as an elegant way to describe pattern matching with an
infinite set of patterns.
For instance, consider the definition of \code{last}
shown above. Since the equational condition requires
the complete evaluation of the input list,
an expression like \code{last$\;$[failed,3]} (where \code{failed}
is an expression that has no value) can not be evaluated
to some value.
Now, consider that \code{last} is defined by the following (infinite)
set of rules:
\begin{curry}
last [x] = x
last [x1,x] = x
last [x1,x2,x] = x
$\vdots$
\end{curry}
Then the expression above is reduced to the value \code{3}
by applying the second rule.
This set of rules can be abbreviated by a single rule:
\equprogram{\label{def:last-fp}%
last (\us{}$\;$++$\;$[x]) = x
}
Since the argument contains the defined operation \ccode{++},
it is called a \emph{functional pattern}.
Conceptually, a functional pattern denotes all constructor terms
to which it can be evaluated (by narrowing).
In this case, these are the patterns shown above.
Operationally, pattern matching with functional patterns
can be implemented by a specific unification procedure
which evaluates the functional pattern in a demand-driven manner
\cite{AntoyHanus05LOPSTR}.
Functional patterns are useful to express pattern matching
at arbitrary depths in a compact manner.
For instance, they can be exploited for a compact and declarative
approach to process XML documents \cite{Hanus11ICLP}.

A delicate point of functional patterns are
non-linear patterns, i.e., if a functional pattern
is evaluated to some constructor term containing multiple
occurrences of a variable.
For instance, consider the function
\begin{curry}
dup :: a -> (a,a)
dup x = (x,x)
\end{curry}
and its use in a functional pattern:
\begin{curry}
whenDup (dup x) = x
\end{curry}
By the semantics of functional patterns, the latter rule is
equivalent to the definition
\begin{curry}
whenDup (x,x) = x
\end{curry}
Due to the non-linear left-hand side,
the type of \code{whenDup} is
\begin{curry}
whenDup :: Data a => (a,a) -> a
\end{curry}
Now, consider the operation \code{const} defined by
\begin{curry}
const :: a -> b -> a
const x _ = x
\end{curry}
and its use in a functional pattern:
\equprogram{\label{def:g-const}%
g (const x x) = x
}
By the semantics of functional pattern, the definition of \code{g} is
equivalent to
\begin{curry}
g x = x
\end{curry}
so that a correct type is
\begin{curry}
g :: a -> a
\end{curry}
Hence, the type context \code{Data$\;$a} is not required,
although the variable \code{x} has a multiple occurrence in (\ref{def:g-const}).
This example shows that, if functional patterns are used,
the requirement for a \code{Data} context depends on the
linearity of the constructor terms to which the
functional patterns evaluate.
Since this property is undecidable in general,
a safe approximation is to add a \code{Data} constraint
to the result type of the functional pattern.
This has the consequence that the type of \code{last},
when defined as in (\ref{def:last-fp}), is inferred as
\begin{curry}
last :: Data a => [a] -> a
\end{curry}
Basically, this type is the same as we would obtain when
defining \code{last} with an
equational constraint, but it could be done better:
since the functional pattern \code{(\us{}$\;$++$\;$[x])}
always yields a linear term, the type class constraint \code{Data$\;$a}
is not necessary.
Hence, one can make the type checking for operations
defined with functional patterns
more powerful by approximating the linearity property
of the functional pattern.
Such an approximation has already been used in \cite{AntoyHanus05LOPSTR}
to improve the efficiency of the unification procedure
for functional patterns.
However, a significant drawback would be the fact that the inferred type
of a function would depend on the quality of the approximation.
As a consequence, the principal type of a function \cite{Hindley69,Damas82}
would become ambiguous under certain circumstances and
would depend on a function's implementation.

\section{Related Work}
\label{sec:related}

We already discussed in the previous sections
some work related to the interpretation and use of equality
in declarative languages.
In the following, we focus on some additional work
related to our proposal.

The necessity to distinguish different equalities in the context of
functional logic programming and to define their exact semantics has been
recognized before.
In \cite{GallegoArias07}, the authors introduce several equality
(and disequality) operations, among others also an operation for strict
equality.
However, no explicit distinction between equality and equivalence is made as
only the former is discussed.
Note also that some of these operations became obsolete
with \cite{AntoyHanus14}.

In \cite{Lux2008}, the author discusses the addition
of Haskell-like overloading to Curry.
In doing so, a new type class \code{Equal} that contains the
unification operation \ccode{=:=} is proposed.
The intent is to restrict this operation similarly to the equivalence
operation \ccode{==} so that it is only applicable to certain types.
In contrast to our proposal,
it is not enforced that instances of the \code{Equal} type class
should always have the same form.
In the same work, another type class \code{Narrowable} containing a method
called \code{narrow} is proposed in order to restrict the type of logical
variables against the background of higher-rank types.
The method \code{narrow} is very similar to our method \code{aValue}.
But aside from a few downsides of the introduction of such a method,
e.g., a possibly fixed order when enumerating solutions,
no further consequences for the language itself are discussed in that work.

The idea to use a type relation to restrict the type of logical variables has
also been introduced in \cite{MehnerEtAl14} for a better characterization of
free theorems.
In \cite{Mehner15Thesis}, a type class \code{Data} is used for the same reason,
but the class is only used as a marker (as in \cite{MehnerEtAl14})
so that the type class does not contain any methods.

On a side note, there is also a \code{Data} type class in Haskell.
However, this particular type class is used for generic programming
in Haskell and shares nothing but the name with our type class
\cite{LaemmelPeytonJones03}.

\section{Conclusions}
\label{sec:concl}

In this paper we presented a solution to various problems
w.r.t.\ equality and logic variables in functional logic programs
by introducing a new type class \code{Data}.
Instances of this class support a generator operation \code{aValue} for values
and a strict equality operation \ccode{===} on these values.
In contrast to other classes, instances of this class
can only be derived in a standard manner and cannot be defined
by the programmer.
This decision ensures a reasonable semantics:
if \code{$e_1\;$===$\;e_2$} evaluates to \code{True},
then the expressions $e_1$ and $e_2$ have an identical value.
Although this is the notion of strict equality proposed for a long time,
Haskell-like overloading of the class \code{Eq} and its operation
\ccode{==} allows to specify that
``some expressions are more equal than others'' \cite{Orwell45}.

At a first glance, it might be unnecessary to add a further
equality operator and base type class to a declarative language.
The advantage is that this supports a clear documentation
for all functions depending on equality,
as it makes a huge difference in functional logic programming whether
one imposes equality or equivalence in a function's implementation.
If a programmer is interested in identical values, she or he has to
use \ccode{===}.\footnote{As discussed in Sect.~\ref{sec:eqopt},
the unification operator \ccode{=:=} does not need to be used
by the programmer since it is an optimization of \ccode{===}.}
If only equivalence is relevant, \ccode{==} is the right choice.
For instance, consider the operation \code{elem} to check
whether an element occurs in a list. The type
\begin{curry}
elem :: Data a => a -> [a] -> Bool
\end{curry}
indicates that this operation succeeds if the element actually occurs
in the list, whereas the type
\begin{curry}
elem :: Eq a => a -> [a] -> Bool
\end{curry}
indicates that it succeeds if some equivalent element is contained
in the list.

Unfortunately, these details are often not taken into account.
As discussed in this paper, many textbooks and program documentations
simply ignore such differences or are not formally precise
in their statements.

We showed that our proposal is also useful
to type logic variables in a more meaningful way.
The type of a logic variable is required to be an instance of \code{Data}
so that one can enumerate the possible values of this variable.
Although logic variables are often instantiated by narrowing
or unification to appropriate values,
there are situations where an explicit enumeration is necessary
to ensure completeness.
For instance, consider the encapsulation of non-deterministic
computations in order to reason about the various outcomes.
Set functions \cite{AntoyHanus09} are a declarative, i.e.,
evaluation-independent, encapsulation approach.
If $f$ is a (unary) function, its set function $\setfun{f}$
returns the set of all results computed by $f$ for a given argument.
For instance, \code{someDup\setfun$\;$xs} returns the set
of all duplicate elements (see Ex.~\ref{ex:concdup})
occurring in the list \code{xs}.
An important property of a set function is that it encapsulates
only the non-determinism caused by the function's definition
and not by the arguments.
Hence, \code{someDup\setfun$\;$([1,1] ? [2])}
yields two different sets: $\{\code{1}\}$ and $\{\}$.
This property of set functions is important to ensure
their declarative semantics.
It has the consequence that arguments
must be evaluated \emph{outside} the set function.
Hence, to evaluate the expression
\begin{curry}
let x free in $\ldots$($f\setfun$ x)$\ldots$
\end{curry}
it is not allowed to bind \code{x} inside the evaluation of $f$.
As a consequence, \code{x} must be instantiated outside
in order to proceed a computation where $f$ demands its argument.
This can easily be obtained by the use of the operation \code{aValue}:
\begin{curry}
let x = aValue in $\ldots$($f\setfun$ x)$\ldots$
\end{curry}
In order to evaluate the practical consequences of our proposal,
we implemented it in a prototypical manner
in our Curry front end that is used by various Curry implementations.
The changes in the type checker were minimal (e.g., adding 
\code{Data} contexts to the inferred types of logic variables).
Concerning libraries, only a single type signature had to be adapted
in the standard prelude, one of the largest Curry modules:
the type of the ``arbitrary value'' operation gets a \code{Data} context:
\begin{curry}
unknown :: Data a => a
unknown = let x free in x
\end{curry}
In other libraries, only a few types (related to
search encapsulation primitives) had to be adapted.
With these few changes, even larger Curry applications
could be compiled without problems.
This demonstrates that our proposal is a viable alternative
to the current unsatisfying handling of equality and logic variables
in Curry.
Usually, no changes are necessary in existing Curry programs.
Only in the rare cases of function definitions with 
polymorphic non-linear left-hand sides or polymorphic logic variables,
type signatures have to be adapted.

\subsubsection{Acknowledgments.}
The authors are grateful to Sandra Dylus and Marius Rasch for fruitful discussions during the conception phase of this paper.
Furthermore, we thank Kai-Oliver Prott for his efforts
in evaluating our proposal by prototypically implementing it.


\end{document}